\documentstyle[spie, epsfig]{article} 
\input{psfig}   

\title{Obervational Model for Microarcsecond Astrometry with \\
the Space Interferometry Mission}
\author{Mark H. Milman%\supit{b} 
~and
Slava G. Turyshev%\supit{a}
\skiplinehalf 
Jet Propulsion Laboratory, California Institute of Technology,
Pasadena, CA 91109
}

\authorinfo{Further author information: (Send correspondence to
S.G.T.)\\
M.H.M.: E-mail: Mark.Milman@jpl.nasa.gov \\
S.G.T.: E-mail: Slava.Turyshev@jpl.nasa.gov}
\begin{document} 
\maketitle 
%********************************************************************
\baselineskip=.33in    
%#%   
%\baselineskip=.175in
%********************************************************************

%\newpage
%%%%%%%%%%%%%%%%%%%%%%%%%%%%%%%%%%%%%%%%%%%%%%%%%%%%%%%%%%%%% 
\begin{abstract}
The Space Interferometry Mission (SIM) is a space-based
long-baseline optical interferometer for precision astrometry.  One of
the primary objectives of the SIM instrument is to accurately determine
the directions to a grid of stars, together with their proper motions and
parallaxes, improving a priori knowledge by nearly three orders of
magnitude. 
The basic astrometric observable of the instrument is the
pathlength delay, a measurement  made  by a combination of internal
metrology measurements that determine the distance the starlight travels
through the two arms of the interferometer and a measurement of the white
light stellar fringe to find the point of equal pathlength.  Because this
operation requires a non--negligible integration time to accurately
measure the stellar fringe position, the interferometer baseline
vector is not stationary over this time period, as its absolute length and
orientation are time--varying.  This conflicts with the
consistency condition necessary for extracting the astrometric
parameters which requires a stationary baseline vector. This paper
addresses how the time-varying baseline is  ``regularized'' so that it
may act as a single baseline vector for multiple stars, and thereby
establishing the fundamental operation of the instrument. 

\end{abstract}

%>>>> Please include a list of keywords after the abstract 


%%%%%%%%%%%%%%%%%%%%%%%%%%%%%%%%%%%%%%%%%%%%%%%%%%%%%%%%%%%%%
%\newpage
%%%%%%%%%%%%%%%%%%%%%%%%%%%%%%%%%%%%%%%%%%%%%%%%%%%%%%%%%%%%% 
\section{INTRODUCTION}
\label{sect:intro}  % \label{} allows reference to this section

SIM is designed as a space-based 10-m
baseline Michelson optical interferometer operating in the visible waveband.
This mission will open up many areas of astrophysics, via astrometry with
unprecedented accuracy.  Over a narrow field of view SIM is expected to
achieve a mission accuracy of 1 $\mu$as.  In this mode  SIM will
search for planetary companions to nearby stars  by detecting the
astrometric ``wobble'' relative to a nearby ($\le 1^\circ$) reference
star.  In its wide-angle mode, SIM will be capable to provide a 4
$\mu$as~ precision absolute position measurements of stars, with
parallaxes to comparable accuracy, at the end of a 5-year mission.  The
expected proper motion accuracy is around 4 $\mu$as/yr, corresponding to
a transverse velocity of 10~m/s at a distance of 1~kpc.\cite{sim}

%%
%\begin{figure}[h]
%\noindent \vskip -20pt 
%\centering\epsfig{file=metr5.eps,width=14cm}\\[-20pt]
%\caption{Overview of SIM conceptual design: external metrology measures
%relative orientation of  science and guide interferometers baselines.
%\label{fig:exter_metr}}
%\end{figure}
%%
%Figure \ref{fig:exter_metr} shows the design for SIM, based on an
%architecture in which there are three separate baseline vectors.  
%Details of the design may be found in \cite{sim}.

The SIM instrument does not directly measure the angular separation
between stars, but the projection of each star direction vector onto the
interferometer baseline by measuring the pathlength delay of starlight as
it passes through the two arms of the interferometer. The delay
measurement is made by a combination of internal metrology measurements
to determine the distance the starlight travels through each arm, and a
measurement of the central white light fringe to determine the point of
equal pathlength. 

SIM surveys the sky in units called {\it tiles}. A tile is defined as a
sequence of measured delays corresponding to multiple objects all made by
a single baseline vector $\vec{b}$ and central pointing of the
instrument -- that is, all the measurements in a tile are from objects
that are within a single astrometric FOR (field of regard of the
instrument), which is approximately $15^\circ \times 15^\circ$. The
existence of a single baseline vector insures that the system of
equations developed from the observations to extract the astrometric
parameters is not underdetermined. However, the collection of such a
measurement set with a single interferometer is actually impossible, as
the data collection on a sequence of objects takes finite time, over
which both the baseline length and orientation do not remain constant. 

This paper describes the fundamental steps of how the on--board
instrumentation of external metrology and auxiliary guide interferometers
are used to reconstruct the baseline vector sufficiently accurately  so
that it can effectively be modeled as a single vector over the period of
a tile observation. This process has been previously referred to as the 
\emph{regularization} of the baseline \cite{Boden}. The notion of the
regularized baseline has been used extensively in a number of grid
simulation studies that plan observation sequences, predict mission
accuracy, and determine sensitivities to various instrument parameters
\cite{Boden,Swartz,LoiseauMalbet}. 

The process of reconstructing the baseline vector when implemented
onboard in real--time  is termed pathlength feedforward, and is a
critical component to the operation of the interferometer. Because many
of the astrometric targets will be very dim, it is not possible for the
science interferometer to track the fringes and compensate for optical
pathlength difference variations in real time using the dim target as 
the signal. As a result the fringes associated with the science target
will be washed out due to uncontrolled motions of the instrument. The
adopted solution in these cases is to use precise attitude information
obtained from the two guide interferometers and construct a delay
tracking signal that will be fed to the science interferometer's delay
line in an open loop fashion. This aspect of the baseline
regularization process will be covered in some detail.  An overview of
the mission and several of the major subsystems of the instrument can
be found in references \cite{Marr,Brady,Stubbs,Bell}

\section{Astrometry with SIM}  
\label{sec:astrom}

SIM is designed to measure the pathlength delay between the two arms of
the interferometer. The instantaneous   delay value is given formally
by the interferometer astrometric equation: \cite{LoiseauMalbet} 
\begin{equation}
d(t) = \big(\vec{b}(t) \cdot \vec{s}\big) + k + \eta(t),
\label{eq:del}
\end{equation}
\noindent where $d$ is the external optical pathlength delay synthesized
by a combination of internal metrology and white light fringe estimation,
$\vec{s}$ is the normal to the wavefront of the starlight (the unit
3-vector to the observed object),
$\vec{b}$ is the baseline 3-vector,   $k$ is a so-called constant (or
calibration) term that represents possible optical path differences
between the light collected from the target object and the internal
metrology, and $\eta(t)$ is the noise in the measurement. 
 
%-------------
\begin{figure}
%\vskip-30pt
\centering\epsfig{file=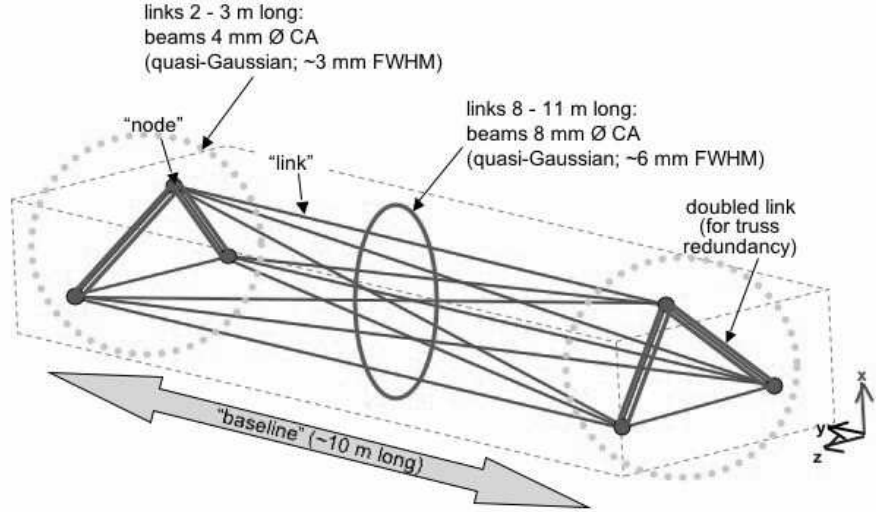,width=14cm}\\[-60pt]
\caption{SIM External metrology design}
\label{fig:sim_design}
\end{figure}
%-------------

Because of limitations imposed by the optical throughput of the system and the
brightness of the observed objects several  seconds of integration time are
necessary to bring the average white light fringe estimate to the required
accuracy. Thus, the actual instrument measurement is the following: 
\begin{equation}
 \langle d_i\rangle=\big(\vec{s}_i\cdot
\langle \vec{b} \rangle\big)+k+\eta_i,
\label{eq:del1}
\end{equation} 
where $\langle d_i\rangle$ is the average measured external delay
obtained from internal metrology measurements and white light fringe
estimation, 
$\langle \vec{b} \rangle$ is the average baseline vector over the period
of the observation, and $\eta_i$ is the measurement noise.  

The fundamental objective of the instrument is to make these delay
measurements so that the astrometric parameters of position, proper motion,
and parallax can be ascertained for the stars that are observed.  Focusing
just on the problem of estimating stellar positions, the astrometry problem
is to determine the vector $\vec s$ in Eq. (\ref{eq:del1}) above.  However, 
{\it all} of the quantities above on the right must be treated as unknown
because of the $\mu$as level precision requirements of the instrument.  (The
Hipparcos catalogue has an accuracy on the order of several mas for stellar
positions, while standard attitude determination and alignment systems on
spacecraft can determine the baseline vector to the order of an arcsec.)  SIM
circumvents this difficulty by observing multiple stars, $\lbrace
s_i\rbrace$, within its field of regard so that the observation equations can
be modeled as

\begin{equation} 
d_i=s_i\cdot b+k+\eta_i,\qquad i=1,...,N.
\label{eq:del2}
\end{equation}
The critical assumption here is that there is a single
baseline vector, albeit unknown, that needs to be solved for as well as a
single constant term.  Thus, Eq. (\ref{eq:del2}) consists of $N$ equations
with $2N+4$ unknowns.  If a new baseline orientation is used to observe
the same set of stars, an additional $N$ equations are obtained, but only
at the expense of 4 additional unknowns necessary to determine the new
baseline vector and constant term.  It is evident with more observations
the system of equations eventually becomes overdetermined so that the
stellar positions can be resolved.   Applying this idea to a set of
tiles that covers the entire celestial sphere is the kernel of SIM's
strategy to perform wide angle astrometry.

The observable in (\ref{eq:del1}) is the average delay measurement made by
the interferometer.  Considerable analysis, simulation and technology
development and validation has been devoted to the problems of precision
white light fringe estimation and metrology that are used to synthesize this
observable.\cite{Laskin,MilmanBasinger,Milman,MCT,Zhao,Shen,Turyshev03}  The
focus of this paper is to show that the model equations Eq. (\ref{eq:del2})
are actually consistent with the data that is collected by the instrument.  To
see where a potential conflict may arise, observe that because the baseline
vector is time--varying, the average position over the integration period
changes from observation to observation.  The resulting model over a tile is

\begin{equation}
\langle d_i\rangle= s_i\cdot \langle b_i\rangle+k,\quad i=1,...,N.
\label{eq:del4}
\end{equation} 
The difficulty now is that there is a different
(unknown) average baseline for each star;  violating the assumption made
in (\ref{eq:del2}).

In principle SIM solves this problem by using guide interferometers and
external metrology to track the baseline vector during the observations. 
The details of this process will be developed over the next two
sections.  Here will give the overview of how this is done, specifically
with respect to overcoming the difficulty posed in (\ref{eq:del4}).

SIM uses two auxiliary guide interferometers that lock on bright ``guide''
stars ($\vec{g}_1$, $\vec{g}_2$), thereby keeping track of the directions
to these stars, and hence also of the  rigid-body motion of the
instrument.   The third interferometer switches between the science
targets ($s_{\tt 1}$, $s_{\tt 2}$, $\cdots$), measuring the projected
angles between the targets and the interferometer baseline vector. An
external metrology system keeps track of the flexible-body motions of the
instrument by measuring changes in the baseline vectors of the three
interferometers in a local frame, or equivalently, determining their
relative orientations. 

Because the three interferometer baselines are not collinear, to complete
the characterization of the rigid body behavior of the instrument a third
inertial measurement is required. This measurement is termed the ``roll''
measurement. In the ideal case we show that the SIM instrumentation is
sufficient in the sense that in the absence of measurement errors and
{\it a priori} parameter errors, the collection of measurements made by
the guide interferometers, the roll measurement, the external metrology
measurements together with the {\it a priori} parameter data consisting of
the positions of the guide stars, the initial guide and science baseline
vectors in the local frame, uniquely determine the baseline vector of the
science interferometer in inertial space.
 
When this is the case the vector $\langle b_i\rangle$ in (\ref{eq:del4})
is known for all $i$.  In reality there are both measurement errors and
{\it a priori} parameter errors so it can never be assumed that the
$\langle b_i\rangle$ are known. For any parameter vector $p$, let $b(t;p)$
be the estimate of the baseline vector using this parameter vector.  In the
absence of measurement error the sufficiency of the SIM instrumentation is
embodied in the statement that $b(t)=b(t;p_0)$ where $p_0$ is the true
parameter vector.  Let $p$ denote the {\it a priori} parameter vector and
write
\begin{equation}
\delta \vec b(t)=\vec b(t;p_0)-\vec b(t;p).
\end{equation} 
In Section \ref{sec:regul} we show that for $|p-p_0|$ 
sufficiently small
\begin{equation}
\delta \vec{b}(t)=\delta \vec{b}^0+\vec \epsilon(t),
\label{eq:delta_p}
\end{equation} 
where $\delta b_0$ is a constant vector (so long as the
guide interferometers are locked on the guide stars) and $|\epsilon(t)|$
is a residual variation that contributes a delay error much less than the
magnitude of the measurement noise $\eta_i$ and can be ignored in the
analysis.

Now we may write the instantaneous delay equation as
\begin{eqnarray} 
d_i(t)
&=& \vec s_i\cdot \vec b(t;p)+\vec s_i\cdot \vec{\delta b}(t)+
k\nonumber\\
&=& \vec s_i\cdot \vec b(t;p)+\vec s_i\cdot \vec{\delta b}^0+ 
\vec s_i\cdot \vec \epsilon(t)+k.
\end{eqnarray} 
Let $\vec s_i{}^0$ denote the {\it a priori} estimate of the
position so that $\vec s_i=\vec s_i{}^0+\vec{\delta s_i}$, where 
$\vec{\delta s_i}$ is the correction vector that is sought.  Thus, after
averaging 
\begin{equation} 
\langle d_i(t)\rangle - \vec  s_i{}^0\cdot \langle \vec b(t;p)\rangle =
\vec {\delta s_i{}^0}\cdot \langle \vec b(t;p)\rangle + \vec s_i{}^0\cdot 
\langle\delta \vec b^0\rangle +\vec s_i{}^0\cdot
\langle \epsilon(t) \rangle + k.
\end{equation} 
The quantity on the left is termed the ``regularized delay'' and is the 
quantity that is synthesized from the SIM instrumentation.  And since
$\vec b(t;p)$ is known and the contribution of the term containing
$\epsilon(t)$ can be ignored, the unknowns are the corrections to the
science star positions and the correction to the baseline vector which is now
a single constant vector over the entire tile.  In this sense the
idealized model equations in (\ref{eq:del2}) are correct.

\subsection{The ideal instrument measurements}

We will now get into the details of how $\langle \vec{b}\rangle$ is
obtained from the measurements and {\it a priori} data.   For this
purpose it suffices to treat SIM as a set of fiducials,
$\vec{X}_1,...,\vec{X}_N,~\vec{X}_i\in{\bf R^3}$.  (Here ${\bf R^3}$
denotes Euclidean 3--space.)  In the SIM reference design, shown in
Figure 1, there are a total of six fiducials, and in a spacecraft local
frame their  coordinates are collected in the matrix $\vec X_{\tt 
fid}$ (with units in meters):
\begin{equation} 
\vec X_{\tt  fid}=
\left[\matrix{0&0&0\cr 1.4&.55&-1.4\cr
0&0&-2.8\cr 0 &10&-2.8\cr 1.4&9.45&-1.4\cr 0&10&0\cr}\right].
\end{equation} 
The $i^{th}$ row of $X_{\tt  fid}$ defines the coordinates of
$\vec{X}_i$.

The science interferometer baseline vector $\vec b_s$ is defined as
\begin{equation} 
\vec b_s= \vec X_{\tt  fid}(6,:)-\vec X_{\tt  fid}(1,:).
\label{eq:b_s}
\end{equation} 
The auxiliary science baseline vector is defined by fiducials 3 and 4. 
The guide interferometer baseline  vector $\vec b_g$ is
defined
\begin{equation} 
\vec b_g=\vec X_{\tt  fid}(5,:)-\vec X_{\tt  fid}(2,:),
\end{equation} 
and the ``roll'' vector is
\begin{equation}
\vec\tau=\vec X_{\tt  fid}(2,:)-\vec X_{\tt  fid}(1,:).
\end{equation}

For any pair of vectors in ${\bf R^3}$, say $\vec{W}_i$ and $\vec{W}_j$,
the vector $\vec{W}_{ij}$ will denote the difference
$\vec{W}_{ij}=\vec{W}_i-\vec{W}_j$. Thus, for example $\vec b_s=
\vec{X}_{16}$. Our interests center around the evolution of the fiducials
$\vec{X}_i(t)$ over a time period
$t_0\le t\le T$, where $t=t_0$ denotes the beginning of an observation of
a tile and $t=T$ is the time of completion. The problem is solved using
the on--board optical sensing systems that include the external metrology
system, the guide star interferometers, and the roll estimator. The
signals from these systems are briefly described next. 

In the SIM reference design relative distance measurements are made
between each pair of fiducials except for the direct link connecting the
active science interferometer fiducials.  The observed variables
associated with the external metrology system are 
\begin{equation}
\ell_{ij}(t)= |\vec{X}_{ij}(t)|-|\vec{X}_{ij}(t_0)|, 
\qquad j>i, i=1,...5, ~~{\rm excluding} ~(i=1 \,\,\&\,\, j=6).
\label{(2.1)} 
\end{equation}
These measurements are relative distance measurements, and (\ref{(2.1)})
is valid for any choice of coordinate frame. Thus, $\vec{b}_{\tt s}$ is
determined in the local frame from (\ref{eq:b_s}). The problem is to
find $\vec{b}_{\tt s}$  in the inertial frame. This connection is made
with the guide interferometers.

SIM uses a pair of guide stars to produce two independent delay
measurements per observation:
\begin{equation}
d_{\tt gA}=(\vec{g}_{\tt A}\cdot\vec{b}_{\tt gA})+k_{\tt gA}, 
\label{(2.2)}
\end{equation}
\noindent where  $\vec{g}_{\tt A}$ is the position vector to guide star
$A$, with $A\in (1,2)$.

The roll estimator produces a ``measurement'' similar to the guide
interferometers. We designate a guide telescope for use as part of the
roll estimation scheme, and let $\vec{n}$ denote the line--of--sight
vector of this telescope. Next we introduce a fiducial rigidly attached to
the telescope which is also measured by the external metrology system.
Let $\vec{\tau}$ denote the vector connecting this fiducial with the
fiducial mounted on the chosen guide telescope. The
rigidity assumption is that 
\begin{equation}
\kappa=(\vec{n}\cdot\vec{\tau})
\label{(2.3)}
\end{equation}
where $\kappa$ is constant for all values
of $\vec{n}$ and $\vec{\tau}$ over the period of a tile observation. 

\subsection{The logic of SIM astrometric observations} 

Now we will describe how (\ref{(2.1)})--(\ref{(2.3)}) are used to
characterize the evolution of $\vec{b}_{\tt s}$ in the inertial frame. 

First let us see what can be learned from the observations in
(\ref{(2.1)}). Set
$\vec{X}=(\vec{X}_1,...,\vec{X}_N)$, and define $F(\vec{X})$ as the
function with components
$F_{ij}(\vec{X})$ 
\begin{equation}
 F_{ij}(\vec{X})= |\vec{X}_{ij}(t)|-|\vec{X}_{ij}(t_0)|,\qquad
j>i; \quad i=1,...5, \quad i\not=1 \quad \& \quad j\not=6.
\label{(2.4) }
\end{equation}
We seek the solution to the system of equations 

\begin{equation}
F(\vec{X})=\ell,\quad \ell=(\ell_{ij}).
\label{(2.5)}
\end{equation}
The first thing to note is that if $\hat{U}$ is a rotation matrix acting
on vectors in ${\bf R^3}$, and if $\vec{\xi}\in {\bf R^3}$, then 
\begin{equation}
 F(\vec{X})=F(\vec{Y}),\quad\hbox{where}\quad
\vec{X}=(\vec{X}_1,...,\vec{X}_N),\quad
\vec{Y}=(\hat{U}\vec{X}_1+\vec{\xi},...,\hat{U}\vec{X}_N+\vec{\xi})\label{(2.6)} 
\end{equation}
since 
\begin{equation}
|\hat{U}\vec{X}_i+\vec{\xi}-(\hat{U}\vec{X}_j+\vec{\xi})|=
|\hat{U}\vec{X}_{ij}|=|\vec{X}_{ij}|.\label{(2.7)}
\end{equation}
The final equality in (\ref{(2.7)}) follows because $\hat{U}$ is
orthogonal, and hence, preserves norms. Importantly the {\it converse} of
(\ref{(2.6)}) also holds:
\vskip 4pt  Property 1: {\it If $F(\vec{X})=F(\vec{Y})$ for some pair
$\vec{X}$ and $\vec{Y}$ (in a small neighborhood $N$ of $\vec X_0$),
then there exists a rotation matrix $\hat{U}$ and 3--vector $\vec{\xi}$
such that
$\vec{Y}_i=\hat{U}\vec{X}_i+\vec{\xi}$ for all $i$.} 
\vskip 4pt  This is the fundamental result which links the local and
inertial frames.  The linear justification of this principle is that
since $F(\vec{X})=F(\vec{Y})$ and to first order
$F(\vec{Y})=F(\vec{X})+F'(\vec{X})(\vec{Y}-\vec{X}),$ it follows
that
$\vec{Y}-\vec{X}$ must be in the kernel of
$F'(\vec{X})$, which can be characterized as the rigid body motions of
the fiducial system. It is not difficult to extend the linear argument to
the full result. 

Let $\vec{X}$ denote the vector of fiducial positions in inertial
coordinates. Then (in the absence of noise) $\vec{X}$ solves (\ref{(2.5)}).
If $\vec{X}^{\tt loc}$ is a solution to (\ref{(2.5)}) computed in a local
spacecraft coordinate frame, then  Property 1 states that there is a rotation
matrix $\hat{U}$ such that 
\begin{equation}
\vec{X}_{ij}=\hat{U}\vec{X}_{ij}^{\tt \, loc}\label{(2.8)} 
\end{equation}
for {\it every} pair of fiducials
$\vec{X}_i$ and $\vec{X}_j$. Thus the matrix $\hat{U}$ is the
transformation between the local and inertial coordinate frames. And
since the science baseline vector $\vec{b}_{\tt s}$ is known in local
coordinates from  external metrology measurements, the   problem  of
determining $\vec{b}_{\tt s}$  in inertial coordinates is 
solved once we obtain $\hat{U}$, viz. 
\begin{equation}
\vec{b}_{\tt s}=\hat{U} \,\vec{b}_{\tt s}^{\tt \, loc}.\label{(2.9)} 
\end{equation}

The equations for obtaining $\hat{U}$ are provided by the guide
interferometer measurements and the roll estimator. The guide
measurements may be written as 
\begin{equation}
d_{\tt gA}=(\vec{g}_{\tt A}\cdot \hat{U} \,\vec{b}_{\tt g}^{\tt \,
loc})+k_{\tt gA},
\label{(2.10)}
\end{equation}
with $\vec{b}_{\tt g}^{\tt \, loc}$ 
assumed known from external metrology data. The third
equation needed to determine $\hat{U}$ is provided by the roll estimator,
which has the form from (\ref{(2.3)})
\begin{equation}
 \kappa =(\vec{n}\cdot\vec{\tau})   
      = (\vec{n}\cdot \hat{U}\vec{\tau}^{{\tt \, loc}}), 
\label{(2.11)}
\end{equation} 
where $\vec{n}$, $\kappa$ and $\vec{\tau}^{\tt loc}$ are all known. We now
have 3 equations with which to determine $\hat{U}$. (The set of orthogonal
matrices live in a space of three dimensions, so three non--redundant
equations are sufficient.) Note that the roll estimator equation has
essentially the same form as the guide equation. The correspondences are
that the guide measurement is replaced by a constant value using the
rigidity assumption of the structure connecting the fiducials used for
roll, the inertial position of the guide star is replaced with the
inertial  line of sight vector of the telescope, and the guide
interferometer baseline vector is replaced with roll vector
connecting the two fiducials. 

\section{Solving the baseline vector equations}
\label{sec:solving}

As described above, there are two components to the problem of determining
$\vec b(t;p)$.  The first part inverts the 1--D external metrology
measurements into position vectors computed in the spacecraft local frame
for all of the fiducials.  The second part uses the guide interferometer
measurements together with the roll estimator equation to determine the
transformation between the local frame and the inertial frame.

\subsection{Inverting the external metrology measurements} Because in
general the system of external metrology equations Eq. (\ref{(2.5)}) is
overdetermined, the estimate of the fiducial positions $\vec{X}_i$ is
derived from the solution to the nonlinear least squares problem
\begin{equation}
\min_X |F(\vec X)-\ell|^2,
\label{eq:24}
\end{equation} 
where $\ell$ is the vector with components $\ell_{ij}$.   In the case of the
reference design, $\ell$ is a 14--vector comprised of links between every
pair of the six fiducials of the external metrology subsystem, except for
the one pair that is a direct link between the fiducials of the active
science interferometer baseline.  Let $F'$ denote the differential of the
function $F$ at $\vec X^0$.  The rows of $F'$ are constructed from the
gradients of the functions
$F_{ij}$.  These gradients are easily calculated analytically as
\begin{equation}
\nabla
F_{ij}(\vec X_0)=[0_{1,3(i-1)},~
\frac{(\vec{X}^0_j-\vec{X}^0_i)^T}{|\vec{X}^0_j-\vec{X}^0_i|},~
0_{1,3(j-i-1)},~
-\frac{(\vec{X}^0_j-\vec{X}^0_i)^T}{|\vec{X}^0_j-\vec{X}^0_i|},~
0_{1,3(N-j)}],
\end{equation} 
where $0_{s,t}$ is the zero matrix with $s$ rows and $t$ columns.  Thus
$F'$ is a $14\times 18$ matrix corresponding to the 14 metrology
measurements and 18 coordinates representing the 6 fiducial positions.
Let $F'^\dag$ denote the pseudoinverse of the differential of
$F$ at $\vec X^0$.  (The pseudoinverse only needs to be calculated once
at the beginning of the tile.)  Then  the iteration scheme beginning with
$\vec X^0=(\vec{X}_1^0,...,\vec{X}_N^0)$,
\begin{equation} 
\vec X^k=\vec X^{k-1}+F'^{\dag}[\ell-F(\vec X^{k-1})],
\label{eq:bigF}
\end{equation}
can be shown to converge to the unique solution of (\ref{eq:24}) using
a Gauss-Newton convergence argument.  This solution is contained in a
ball of radius
$2|F^{\dag}l|$ about $\vec X_0$ and an $O(|\ell|^3)$ error is incurred if the
iteration is stopped at $k=2$; which is  sufficient for both SIM
requirements as the constraints on the magnitude of the flexible body
motion of the fiducials leads to $|\ell|\approx 10^{-5}$m.  The use of
the pseudoinverse in  (\ref{eq:bigF}) is equivalent to constraining
certain linear combinations of fiducial positions to remove the rigid
body motions of the system.  

\subsection{The attitude equations}  

The second part of the algorithm for estimating $\vec b(t)$ requires
solving for the attitude. The three equations for  obtaining the
transformation, $\hat U$, between the local and inertial frames are provided by the guide
interferometer and the roll estimator equations.

$\hat U$ can be parameterized in several ways.  We will make use of the
fact that there is a skew--symmetric matrix $\hat S$ such that $\hat
U=\exp(\hat S)$ so that $\hat U$ has the series expansion
\begin{equation}
\hat U=\hat I+\hat S+\hat S^2/2!+\hat S^3/3!+....
\label{eq:u_hat}
\end{equation} 
We will also use the 1--1 correspondence between the set of skew symmetric
matrices and 3--vectors via the  mapping $\vec{\omega}\rightarrow
\hat S(\vec \omega)$,
\begin{equation} 
\hat S(\vec{\omega})= \left[\matrix{0 & -\omega_3 &\omega_2\cr
                    \omega_3 & 0 & -\omega_1\cr
                       -\omega_2 & \omega_1 &0\cr}\right],\quad
\vec{\omega}=(\omega_1,~\omega_2,~\omega_3).
\label{eq:s_hat}
\end{equation} 
Without loss of generality we may assume that $\hat
U\approx \hat I$ because of on--board attitude knowledge.  The quadratic
approximation to (\ref{(2.10)})--(\ref{(2.11)}) using (\ref{eq:u_hat})
and (\ref{eq:s_hat}) is
\begin{equation}
T_{\tt ext}\vec{\omega}+G_{\tt ext}(\vec{\omega})=y,
\label{eq:tgy}
\end{equation}
where,
\begin{equation}
T_{\tt ext}=
\left[\matrix{\vec{\tau}^{\tt loc}\times \vec{n}\cr T\cr}\right],
\qquad 
T=\left[\matrix{\vec{b}_g^{\tt loc}\times \vec{g}_1\cr
            \vec{b}_g^{\tt loc}\times \vec{g}_2\cr}\right],
\end{equation}
\begin{equation}
G_{\tt ext}(\vec{\omega})=
\left[\matrix{0\cr G(\vec{\omega})\cr}\right], 
\qquad 
G(\vec{\omega})=-{1\over 2}\left[\matrix{\langle \vec{\omega}\times
\vec{g}_1,\vec{\omega}\times \vec{b}_g^{\tt loc}\rangle\cr
            \langle \vec{\omega}\times \vec{g}_2,\vec{\omega}\times
\vec{b}_g^{\tt loc}\rangle\cr}\right],
\label{eq:big_g}
\end{equation} 
and
\begin{equation} 
y=\left[\matrix{\kappa-\vec{n}\cdot\vec{\tau}^{\tt loc} \cr
  d_{g1}- \vec{g}_1\cdot\vec{b}_g^{\tt loc}\cr
  d_{g2}- \vec{g}_2\cdot\vec{b}_g^{\tt loc}\cr
          }\right].
\label{eq:y_martix}
\end{equation} 
It can be shown that the solution to (\ref{eq:tgy})  produces an error of
$O(|\vec{\omega}|^4)$ in determining $\hat U$. (This is less than a
$10^{-15}$ rad error for 20 arcsec of motion of the instrument.)  Without
going into the details of the proof of this result here, we just remark
that it hinges on the simple observation that for any skew symmetric
matrix $\hat S$, the distance from $\hat I+\hat S+\hat S^2/2$ to an orthogonal
matrix is
$|O(\hat S^4)|$.  And this follows by noting that
\begin{eqnarray} 
(\hat I+\hat S+\hat S^2/2)(I+\hat S+\hat S^2/2)^T
&=&\hat I+\hat S+\hat S^T+\hat S\hat S^T+\hat S^2/2+(\hat S^2)^T/2+\nonumber\\
& &+~\hat S(\hat S^2)^T/2+\hat S^2\hat S^T/2+
\hat S^2(\hat S^2)^T/4=\nonumber\\
                       &=&I+\hat S^4/4,
\end{eqnarray}
where $\hat I$ denotes the identity matrix and we have used 
$\hat S^T=-\hat S$ and $(\hat S^2)^T=\hat S^2$.  (Recall that a matrix
$\hat X$ is orthogonal if $\hat X\hat X^T=\hat I$.)

Rewriting (\ref{eq:tgy}) as
\begin{equation}
\vec{\omega}=T_{\tt ext}^{-1}y-T_{\tt ext}^{-1}G_{\tt
ext}(\vec{\omega}),
\label{eq:omega}
\end{equation} 
the solution can be obtained by a fixed point iteration on the mapping
defined above on the right:
\begin{equation}
\vec{\omega}^k=
T_{\tt ext}^{-1}y-T_{\tt ext}^{-1}G_{\tt ext}(\vec{\omega}^{k-1}),
\quad \vec{\omega}_0=T_{\tt ext}^{-1}y;
\label{eq:36}
\end{equation}
with error estimate
\begin{equation} 
|\vec{\omega}^k-\vec{\omega}|\le
|T_{\tt ext}^{-1}y|^{k+1}.
\end{equation} 
A standard contraction mapping argument can be used to establish this
result. The estimated baseline in inertial space is then realized as
\begin{equation}
\vec{b}_s=\vec{b}_s^{\tt loc}+\vec{\omega}\times \vec{b}_s^{\tt loc}
+{1\over2}\vec{\omega}\times (\vec{\omega}\times \vec{b}_s^{\tt loc}),
\end{equation}
where $\vec{b}_s^{\tt loc}$ is determined from (\ref{eq:bigF}), and
$\vec{\omega}$ is obtained from the iteration in (\ref{eq:36}).

To simplify matters a little for the error analysis that will be
performed in Section \ref{sec:regul}, we define $\vec{\omega}_I$ as the
starting value in (\ref{eq:36}):
\begin{equation}
\vec{\omega}_I=T_{\tt ext}^{-1}y,
\label{eq:om1}
\end{equation} 
and $\vec{\omega}_{II}$ as the first iterate:
\begin{equation}
\vec{\omega}_{II}=\vec{\omega}_I-
T_{\tt ext}^{-1}G_{\tt ext}(\vec{\omega_I}).
\label{eq:om2}
\end{equation} 
Because our estimates will carry through second order, the science
baseline estimate we use is
\begin{equation}
 \vec{b}_s=\vec{b}_s^{\tt loc}+\vec{\omega}_{II}\times
\vec{b}_s^{\tt loc}+{1\over 2}\vec{\omega}_I\times (\vec{\omega}_I\times
\vec{b}_s^{\tt loc}).
\end{equation}

\begin{figure}
%\vskip-60pt
\centering\epsfig{file=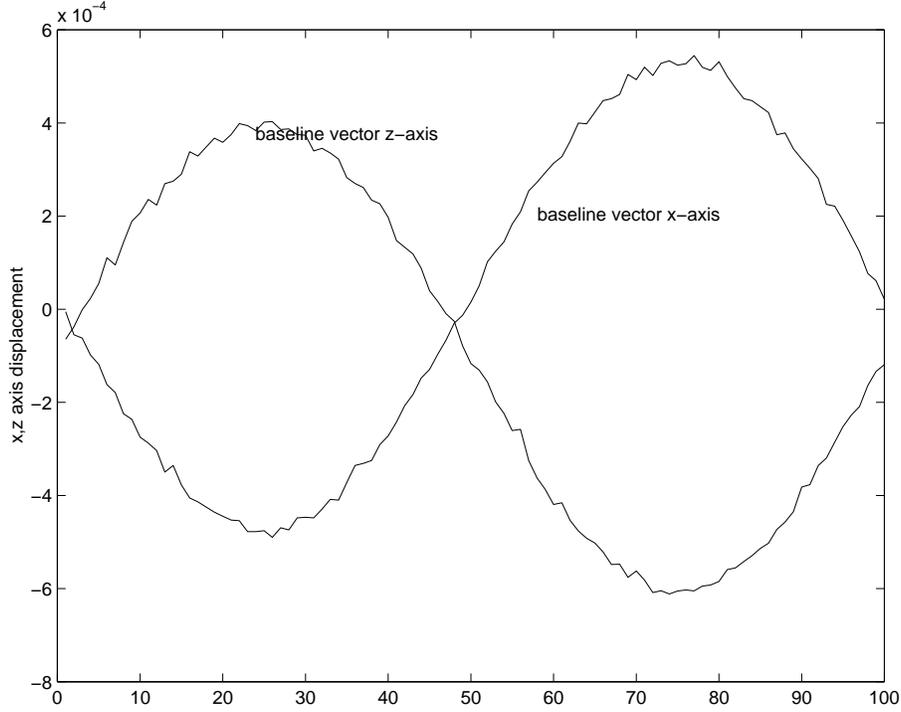,width=12cm}\\%[-70pt]
\caption{x and z baseline displacement vector
components due to rigid body motion}
\label{fig:flex_body}
\end{figure}

\begin{figure}
%\vskip-60pt
\centering\epsfig{file=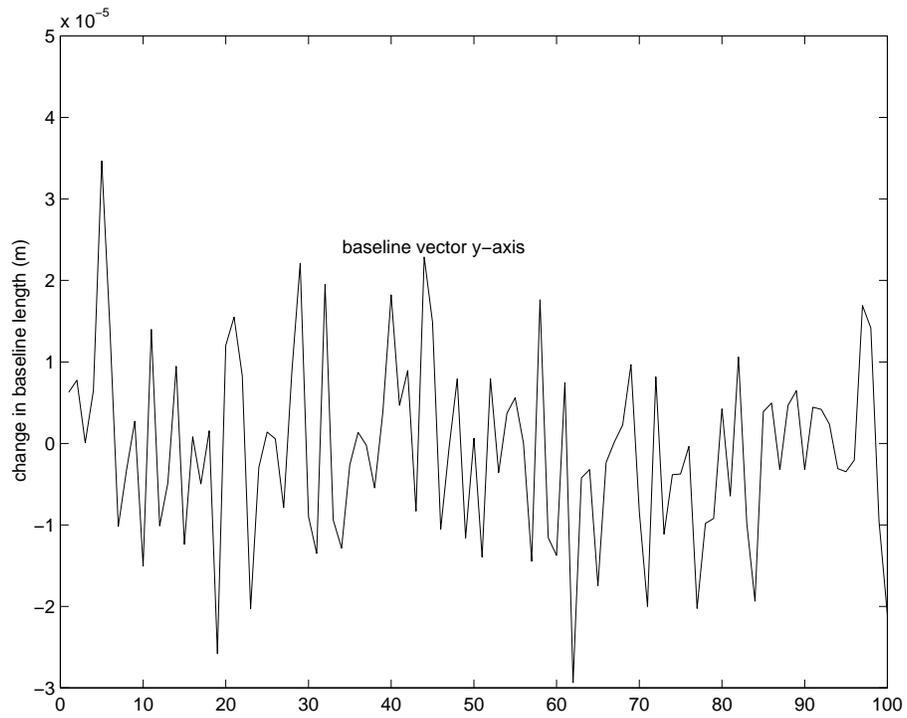,width=12cm}\\%[-70pt]
\caption{y baseline displacement vector component due
to flex body motion}
\label{fig:rig_body}
\end{figure}

\begin{figure}
%\vskip-60pt
\centering\epsfig{file=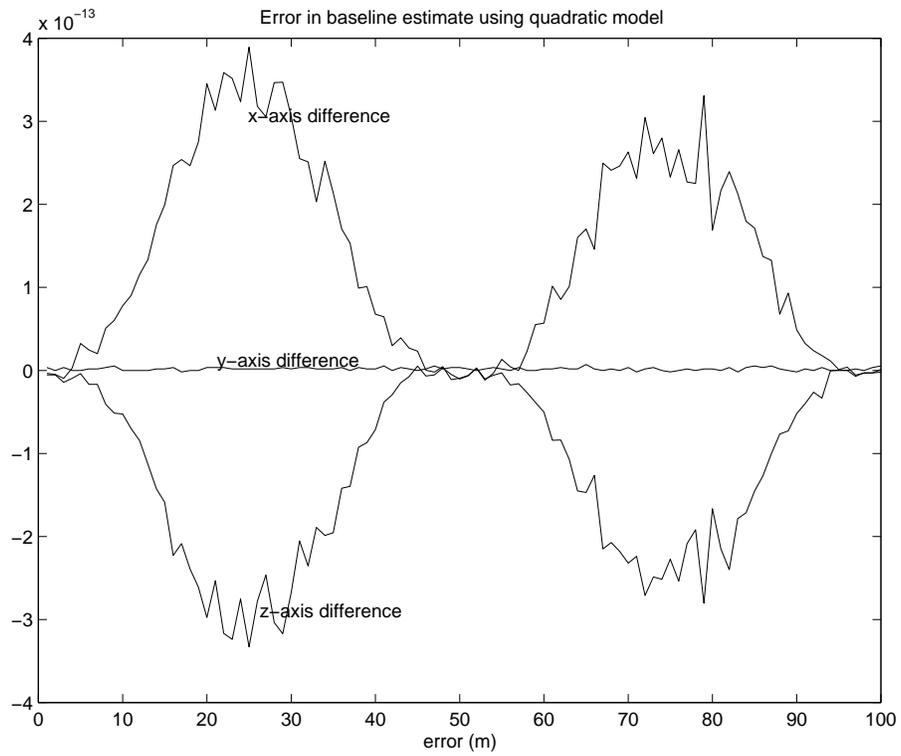,width=12cm}\\%[-70pt]
\caption{Error in baseline vector estimate due to
quadratic approximation}
\label{fig:quatratic}
\end{figure}

This second order approximation accommodates flexible fiducial motions
on the order of tens of microns in the inversion of the external
measurements and rigid body motions of 100 $\mu$rad in the attitude
equations.  These are both significantly larger than SIM requirements. 
As an example of these approximations, the plots in Figures
\ref{fig:flex_body} and \ref{fig:rig_body} show induced motion of this
magnitude in the system.  The $x$ and $z$ axis motion shown are due to
large (20 as) sinusoid motion of the baseline, while the $y$ axis
motion is due mostly to flexible body motion superimposed on the rigid
body displacement. Figure \ref{fig:quatratic} shows the error in the
estimated baseline components using the quadratic model.  It is seen
that these errors are sub--picometer, as the error analysis predicts. 
These methods could of course accommodate a larger range of motion at
the expense of utilizing more iterations in the solution of the
external metrology and attitude equations. A mission level simulation
capability that incorporates these equations and a number of instrument
models is under development.\cite{simsim}

%-------------

%-------------

%-------------

\section{A Geometric Interpretation}
\label{sect:mod}

The solution to the set of non-linear equations that governs the
evolution of the baseline vector has a geometric interpretation that will
be discussed here.  To facilitate this we introduce the baseline strain
parameter $\epsilon$ 
  
\begin{equation}
\epsilon(t)=\frac{ b(t)-b(t_0) }{b(t_0)} 
\label{eq:42}
\end{equation}
which we expect to be $\sim \mu$rad because the change in the baseline
length will be on the order of 10 $\mu$m.

Again retaining terms through second order in $\vec{\omega}(t)$, we
parameterize the instantaneous interferometer baseline, $\vec{b}(t)$,  as
follows: 
\begin{equation}
\vec{b}(t)=b(t_0)\Big(1+\epsilon(t)\Big)
\Big(\vec{n}(t_0)+
[\vec{\omega}(t)\times\vec{n}(t_0)]
+\frac{1}{2!}[\vec{\omega}(t)\times[\vec{\omega}(t)\times\vec{n}(t_0)]]
+{\cal O}\big(\omega^3\big)\Big), 
  \hskip 20pt
\label{eq:base0}
\end{equation}
\noindent where  $b(t_0)$, $\vec{n}(t_0)$ are the initial
baseline length and orientations of the  guide interferometer;
$\epsilon(t)=\Delta b(t)/b(t_0)$ is the
time-varying readings of the external metrology; 
$\vec{\omega}(t)$ is the vector of small attitude changes in
the baseline orientation for the interferometer,
$\vec{\omega}(t_0)\equiv0;$ and $[\vec{a}\times\vec{b}]$ denotes the
cross vector product of vectors $\vec{a}$ and $\vec{b}$.

\subsection{Science  Interferometer  Delay Equation}
%\subsection{Equation for the attitude vector $\vec{\omega}_{\tt s}(t)$}
%\subsection{Expressions for the interferometric delays}

The expression (\ref{eq:base0}) allows us to write the delays for all
three interferometers. Thus, to second order in
$\vec{\omega}$ one obtains the following expressions for the
time-varying delays  of the science and the two guide interferometers: 
\begin{eqnarray}
d_{\tt s}(t)&\equiv&k_{\tt s}(t)+b_{\tt s}(t_0)
\Big(1+\epsilon_{\tt s}(t)\Big)\Big(\vec{s}\cdot \Big\{
\vec{n}_{\tt s}(t_0)+
[\vec{\omega}_{\tt s}(t)\times\vec{n}_{\tt s}(t_0)] +
%\nonumber\\
%&& \hskip 120pt +~~
 \frac{1}{2!}[\vec{\omega}_{\tt s}(t)
\times[\vec{\omega}_{\tt s}(t)\times\vec{n}_{\tt s}(t_0)]] 
+{\cal O}\big(\omega^3\big)\Big\}
\Big),
\label{eq:bs1}\\
d_{\tt g}(t)&\equiv&k_{\tt g}(t)+b_{\tt g}(t_0)
\Big(1+\epsilon_{\tt g}(t)\Big)\Big(\vec{g}_{\tt A}\cdot 
\Big\{\vec{n}_{\tt g}(t_0)+
[\vec{\omega}_{\tt g}(t)\times\vec{n}_{\tt g}(t_0)] +
%\nonumber\\
%&& \hskip 120pt +~~
\frac{1}{2!}[\vec{\omega}_{\tt g}(t)
\times[\vec{\omega}_{\tt g}(t)\times\vec{n}_{\tt g}(t_0)]] 
+{\cal O}\big(\omega^3\big)\Big\}
\Big). \hskip 12pt 
\label{eq:bg1}
\end{eqnarray}

Taking into account that, because of the flexible body motions, 
the rate of the attitude drifts of the guide interferometers is
different from that of the science interferometer, we may write:
\begin{equation}
\vec{\omega}_{\tt g}(t)=\vec{\omega}_{\tt s}(t)+
[\vec{\theta}_{\tt g}(t)\times\vec{\omega}_{\tt s}(t)]+
{\cal O}(\theta^2_{\tt g}),
\label{eq:omegas}
\end{equation}
\noindent where $\vec{\theta}_{\tt g}(t)$  is the rate of temporal drift of
the guide interferometer baseline's orientation  relative to that of  the
science interferometer. Also, due to the flexible body motions, the
interferometers may be misaligned at the beginning of the observations. We
define the contribution of this misalignment vector,  
$\vec{\omega}_{\tt 0g}(t_0)$, to the initial baseline  orientation of the
guide interferometer as  below:
\begin{equation}
\vec{n}_{\tt g}(t_0)=\vec{n}_{\tt s}(t_0)+
[\vec{\omega}_{\tt 0g}(t_0)\times\vec{n}_{\tt s}(t_0)]+
{\cal O}\big(\omega_{\tt 0g}^2\big).
\label{eq:ng1}
\end{equation}

Subtracting the initial conditions from the time-varying delays tracked
by the guide interferometers, Eq.(\ref{eq:bg1}), together with 
Eqs.(\ref{eq:omegas}),(\ref{eq:ng1}), define the instantaneous change in the
science interferometer baseline orientation. However, the obtained  system
of  equations is underdetermined. This is why only two out of three
components of the attitude drift vector  $\vec{\omega}_{\tt s}(t)$ may be
determined this way. We call the undetermined component - the  roll component
- and denote it $\alpha_{\tt s}(t)$. For the determination of this component
of the attitude vector SIM uses the roll estimator given by Eq.(\ref{(2.11)}).

In Section \ref{sec:solving} we developed the iterations for solving the
baseline estimation problem. Here we will present the final solution
for the science interferometer in a form that is amenable to
geometric interpretation. Let us first  define  the
following notations: 
\begin{eqnarray} 
\vec{b}(t_0)= b_{\tt 0}\vec{n}_0&-& {\rm  initial ~science ~baseline
~estimate ~(length,} ~b_{\tt 0}=b_{\tt s}(t_0),  {\rm ~and ~orientation,}
~\vec{n}_0= \vec{n}_{\tt s}(t_0){\rm );}\nonumber\\ 
\vec{s}_j, ~~\vec{g}_{\tt 1},\vec{g}_{\tt 2} &-& {\rm unit ~vectors ~for
~{\it j}^{th} ~science ~star ~and  ~that  
~for  ~the ~two ~guide ~stars;}\nonumber\\
%
%\vec{g}_{\tt 1},\vec{g}_{\tt 2} &-& {\rm  unit ~vectors ~for 
%~the ~two ~guide~stars; }\nonumber\\
%
  d_{\tt s}(t),    d_{\tt g}(t)   &-& {\rm instantaneous 
~interferometric ~delays, ~ of ~science ~ and ~guide ~
interfereometers;}\nonumber\\
  k_{\tt 0s}    &-& {\rm constant ~part ~of
~calibration ~term ~for ~the  ~science ~interferometer ~at ~the
~beginning;}\nonumber\\
  \Delta k_{\tt s}(t), ~\Delta k_{\tt g}(t)  &-& {\rm contribution
~of ~a ~temporal ~drifts ~in ~the ~calibration ~terms;}\nonumber\\
 \vec{\omega}_{\tt 0g} &-& {\rm initial ~ misalignment
~of ~the ~guide ~interferometer ~baseline's  ~orientation 
~relative ~to ~that} 
\nonumber\\ && {\rm   of  ~the ~science ~interferometer;}\nonumber\\
\vec{\theta}_{\tt g}(t) 
&-& {\rm contribution ~of ~the ~temporal ~ drift ~of ~the ~guide
~interferometer's  ~orientation ~relative}  \nonumber\\
&& {\rm  to ~that ~of ~the ~science ~interferometer 
~(taken ~care ~by ~the ~external ~metrology);}
\nonumber\\[0pt]
\alpha_{\tt s}(t) &-& {\rm magnitude ~of ~roll ~of ~the  
~science ~interferometer ~(determined ~from~   
Eq.(\ref{(2.11)})).}   \nonumber 
\end{eqnarray}
Given the solution of the attitude matrix, $\vec{\omega}_{\tt s}(t)$, and
equation (\ref{eq:bs1}), and the presented  notations, the instantaneous
delay of the science interferometer is obtained  in the following form: 
 {}
\begin{eqnarray} 
 d_{\tt s}(t)&=&   k_{\tt 0s}  + \Delta k_{\tt s}(t)  +b_0
\Big(1 + \epsilon_{\tt s}(t)\Big)\bigg\{
\big(\vec{n}_0\!\cdot\!\vec{s}_j\big)+
\Delta z_{\tt ff}(t) \bigg\},
\label{eq:delay} 
\end{eqnarray}
\noindent with the {\it feed-forward} signal, $\Delta z_{\tt ff}(t)$, given
by 
\begin{eqnarray} 
\Delta z_{\tt ff}(t)&=& - \bigg( p_{\tt g1}(t)   \,
\frac{\big(\vec{n}_0\!\cdot\!\big[\vec{g}_{\tt 2}\times\vec{s}_j\big]\big)}
{\big(\vec{n}_0\!\cdot\!\big[\vec{g}_{\tt 1}\times\vec{g}_{\tt 2}\big]\big)}-
  p_{\tt g2}(t)    \,
\frac{\big(\vec{n}_0\!\cdot\!\big[\vec{g}_{\tt 1}\times\vec{s}_j\big]\big)}
{\big(\vec{n}_0\!\cdot\!\big[\vec{g}_{\tt 1}\times\vec{g}_{\tt 2}\big]\big)}
\bigg)-
\nonumber\\[5pt] 
&-&  
\frac{1}{2}
\bigg( p_{\tt g1}(t)  \,
\big[\vec{n}_0\times\vec{g}_{\tt 2}\big]-
 p_{\tt g2}(t)  \,
\big[\vec{n}_0\times\vec{g}_{\tt 1}\big]
\bigg)^{ \hskip -2pt 2}   \!\! 
\frac{\big(\vec{s}_j\!\cdot\!\big[\vec{g}_{\tt 1}\times\vec{g}_{\tt 2}\big]\big)}
{\big(\vec{n}_0\!\cdot\!\big[\vec{g}_{\tt 1}\times\vec{g}_{\tt 2}\big]\big)^3}+
\nonumber\\[5pt] 
&+&   
  \epsilon_{\tt g}(t)   \bigg(
\big(\vec{n}_0\!\cdot\!\big[\vec{g}_{\tt 1}\times
 \vec{\omega}_{\tt 0g} \big]\big)
\frac{\big(\vec{n}_0\!\cdot\!\big[\vec{g}_{\tt 2}
\times\vec{s}_j\big]\big)}
{\big(\vec{n}_0\!\cdot\!\big[\vec{g}_{\tt 1}\times
\vec{g}_{\tt 2}\big]\big)}-
\big(\vec{n}_0\!\cdot\!\big[\vec{g}_{\tt 2}\times
 \vec{\omega}_{\tt 0g} \big]\big)
\frac{\big(\vec{n}_0\!\cdot\!\big[\vec{g}_{\tt1}\times
\vec{s}_j\big]\big)} {\big(\vec{n}_0\!\cdot\!\big[\vec{g}_{\tt
1}\times
\vec{g}_{\tt 2}\big]\big)}\bigg)-\nonumber\\[5pt] 
&-& \bigg( 
  p_{\tt g1}(t)   \,
\big[\vec{n}_0\times\vec{g}_{\tt 2}\big]-
  p_{\tt g2}(t)   \,
\big[\vec{n}_0\times\vec{g}_{\tt 1}\big]\bigg)\times
\nonumber\\[0pt] 
&&\hskip 48pt \times~~\,
\bigg(
\Big( \big(\vec{n}_0\!\cdot\!\vec{g}_{\tt1}\big)\,
\big[\vec{n}_0\times \vec{\omega}_{\tt 0g} \big]-
\Big(\vec{n}_0\!\cdot\!
 \vec{\theta}_{\tt g}(t)\Big)
\,\big[\vec{n}_0\times \vec{g}_{\tt 1} \big]\Big)
\frac{\big(\vec{n}_0\!\cdot\!\big[\vec{g}_{\tt 2}\times
\vec{s}_j\big]\big)}
{\big(\vec{n}_0\!\cdot\!\big[\vec{g}_{\tt 1}\times
\vec{g}_{\tt 2}\big]\big)^2}-\nonumber\\[0pt] 
&&\hskip 56pt- ~~
\Big(\big(\vec{n}_0\!\cdot\!\vec{g}_{\tt
2}\big)\,\big[\vec{n}_0\times
 \vec{\omega}_{\tt 0g} \big]-
\Big(\vec{n}_0\!\cdot\!
 \vec{\theta}_{\tt g}(t)\Big)
\,\big[\vec{n}_0\times \vec{g}_{\tt 2} \big]\Big)
\frac{\big(\vec{n}_0\!\cdot\!\big[\vec{g}_{\tt 1}\times
\vec{s}_j\big]\big)}
{\big(\vec{n}_0\!\cdot\!\big[\vec{g}_{\tt 1}\times
\vec{g}_{\tt 2}\big]\big)^2}\bigg) +\nonumber\\[-2pt] 
&+&
 \alpha_{\tt s}(t) \big[\vec{n}_0\times \big(
 \vec{\omega}_{\tt 0g} -
 \vec{\theta}_{\tt g}(t)\big)\big]  \,
\bigg(
\big[\vec{n}_0\times\vec{g}_{\tt 1}\big]
\frac{\big(\vec{n}_0\!\cdot\!\big[\vec{g}_{\tt 2}\times
\vec{s}_j\big]\big)}
{\big(\vec{n}_0\!\cdot\!\big[\vec{g}_{\tt 1}\times
\vec{g}_{\tt 2}\big]\big)}-
\big[\vec{n}_0\times\vec{g}_{\tt 2}\big]
\frac{\big(\vec{n}_0\!\cdot\!\big[\vec{g}_{\tt 1}\times
\vec{s}_j\big]\big)}
{\big(\vec{n}_0\!\cdot\!\big[\vec{g}_{\tt 1}\times
\vec{g}_{\tt 2}\big]\big)} \bigg)+\nonumber\\[-2pt] 
&+&{\cal O}\Big(\omega^{[3]}_{\tt s}(t); 
\omega^2_{\tt 0};  \epsilon^2_{\tt g}(t);
\theta^2_{\tt g}(t)\Big),
%{\cal O}\Big({\rm third ~order~terms}\Big), 
\label{eq:delay2} 
\end{eqnarray}
  
\noindent where the quantity $ p_{\tt g}(t)$  characterizes  the 
pathlength feed-forward  signal (instrumental drifts) and has the 
form:
\begin{equation}
p_{\tt g}(t) = \Big(1 - \epsilon_{\tt g}(t)\Big) 
\frac{\Delta d_{\tt g}(t)-\Delta k_{\tt g}(t)}
{b_{\tt g}(t_0)}-\epsilon_{\tt g}(t)
\big(\vec{n}_0\!\cdot\!\vec{g}_{\tt A}\big).
\label{eq:delay3}
\end{equation}

A variation in the
average calibration term, $k$, within a tile has implications on the overall
astrometric accuracy. Such inner-tile $k$ variation translates to a
distortion of the relative geometry for the objects in the tile. The
implications of inner-tile $k$ variation on the grid reduction accuracy
is current topic of study.
 
\subsection{Geometric Interpretation}

The expressions for the science delay (\ref{eq:delay})-(\ref{eq:delay3})
obtained above have a clear geometric interpretation. A discussion of 
the significance of the different terms contributing to the
science delay is taken up below. 

The first two terms in (\ref{eq:delay}) are the constant delay
offset and the total temporal variation in this delay since the
beginning of the tile observation. In the current  SIM reference
design,\cite{Marr} the challenge is to keep the temporal variation in this
term small, such that its total contribution to the science
delay will be negligible (on the order of tens of picometers). The terms in
the big parentheses are scaled with the change of the baseline's length,
$\epsilon (t)$; the magnitude of this scaling function is on the
order of $1~ \mu$rad, (c.f. (\ref{eq:42})). The first term in the parenthesis is
the usual form for the interferometric delay $(\vec b\cdot\vec s)$,
which represents the constant delay at the beginning of the observations
of the tile. The second term in this parenthesis is the feed-forward
signal, $\Delta z_{\tt ff}(t)$, which is given in (\ref{eq:delay2}). 
The first two terms in Eq. (\ref{eq:delay2}) are the first- and
second-order feed-forward signals that update the instantaneous
attitude of the science interferometer's baseline vector as measured in
the local inertial frame. The third term in this expression is the
second order contribution introduced by the temporal changes in the
baseline lengths of the two guide interferometers ($\epsilon_{\tt g},
\epsilon_{\tt g} $) which are coupled to their initial misalignments
($\omega_{\tt 0g}, \omega_{\tt 0g}$) with respect to the science
interferometer's baseline vector. The next term in Eq. (\ref{eq:delay2}) is
due to the interaction between the first order feed-forward signal and the
rigid-body motions of the entire extended structure. It's instantaneous
contribution is given by the accumulated temporal drifts in the guide 
interferometers baselines orientation over the time  ($\theta_{\tt
g},\theta_{\tt g}$). The last term -- is the second order contribution of the
science  interferometer's baseline roll vector (given by
$\alpha_{\tt s}$) to the total science interferometer's delay.

One of the results that immediately follows from the presented
analysis is the criteria for choosing the positions of the guide stars
in the tile. Indeed, one may see that all the terms responsible for the
feed-forward signal have the quantity 
$\big(\vec{n}_0\!\cdot\!\big[\vec{g}_{\tt 1}\times
\vec{g}_{\tt 2}\big]\big)$ in the denominator, see Eq. (\ref{eq:delay2}).
This quantity is proportional to the volume of  the tetrahedron constructed
from the three vectors, namely the unit vector of baseline orientation
$\vec{n}_0$ and the directions to the two guide stars $\vec{g}_{\tt 1}$ and
$\vec{g}_{\tt 2}$. As this quantity appears in the denominator, it is
preferable to have it as large as possible, thus minimizing the first and
second order terms. In addition, by minimizing their contributions, one
minimizes the contribution of guide star position errors to the estimate of
science delay. The SIM reference design is near optimal in the sense that the
guide stars are placed 90$^\circ$ apart, with each guide star nearly
orthogonal to the interferometer baseline vectors.

\section{Baseline regularization concept}
\label{sec:regul}
When all the parameters used to determine the baseline vector via the
prescription developed in Section \ref{sec:solving} are without error,
then the estimate of the baseline vector is constructed without error.
But this can never be the case, and in Section \ref{sec:astrom} we
showed that it is sufficient that the error be constant. This is the
linchpin idea for the operation of the instrument.

%-------------
\begin{figure}
%\vskip-60pt
\centering\epsfig{file=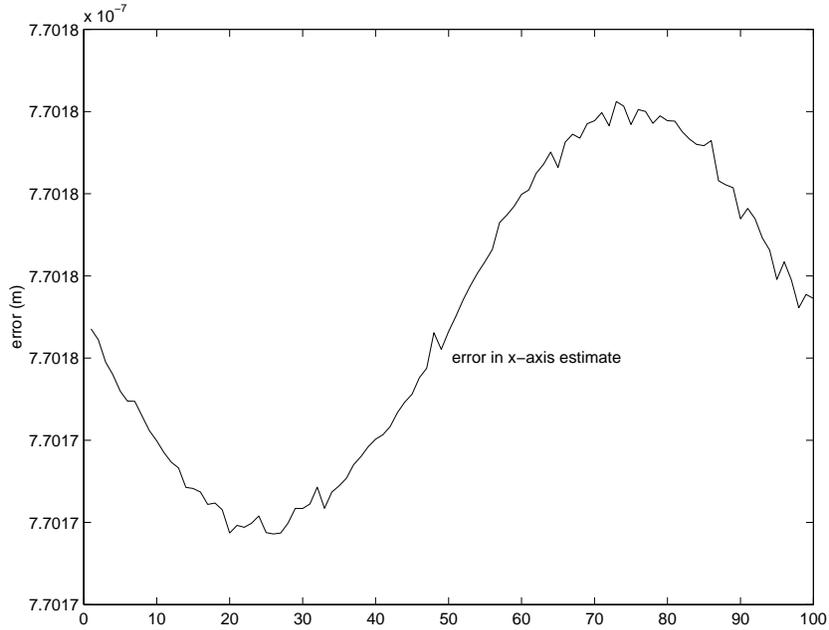,width=11cm}\\%[-70pt]
\caption{  Error in baseline vector estimate due to
guide star position error}
\label{fig:fig3.1}
\end{figure}
%*******************************************************

In Figure \ref{fig:fig3.1} the effect of 20 mas errors in the guide star
directions are introduced in determining the baseline vector. The guide
star position error results in an error in the estimate of the attitude
motion of the system, leading in turn to an error in the baseline
vector estimate.  The $x$ axis error is observed to be on the order of
.0.77 $\mu$m.  However, observe also that this value is highly stable
with a peak to peak range of approximately 10 pm.  Thus although
$\vec b_s(t)- \vec b_s(t;p)$ exhibits micron class error, the important
aspect is that the error is very stable as required in (\ref{eq:delta_p}).

This example provides corroboration of the notion of the regularization
concept introduced in Section 2.  Now we proceed  to give a more
rigorous mathematical justification of regularization in the face of
{\it a priori} parameter errors, such as the guide star position error
discussed above.  

Returning to the prescription for determining $\vec b(t;p)$ in Section
\ref{sec:solving}, {\it a priori} values must be provided for the
following parameters:

\noindent{\bf Guide Interferometer Parameters:}
\begin{itemize}
\item[] $\vec g_i=$ guide star position vectors ($i=1,2)$
\item[] $k_i=$ guide interferometer constant terms ($i=1,2)$
\end{itemize}

\noindent {\bf  Roll Estimator Parameters:}
\begin{itemize}
\item[]
$\kappa=$ inner--product of roll vector $\tau$ with guide telescope 
unit line--of--sight vector
\end{itemize}

\noindent {\bf  Optical Truss Parameters:}
\begin{itemize}
\item[]
$\vec X^0=$ initial vector of all fiducial positions in a s/c local frame
\end{itemize}

The totality of these parameters constitute the parameter vector $p$,
\begin{equation}
 p=(\vec g_1,\vec g_2,c_1,c_2,\kappa,\vec X^0)
\end{equation} 
that is used to define the baseline estimate.  In what
follows we will assume the true parameter vector is
\begin{equation} 
p_0=(\vec g_1+\vec \delta g_1, \vec g_2+ 
\vec{\delta g_2},c_1,c_2,\kappa+\delta\kappa, \vec X^0+\vec{\delta X^0}),
\end{equation} 
while the nominal parameter vector is
\begin{equation} 
p=(\vec g_1,\vec g_2,0,0,\kappa,\vec X^0).
\end{equation} 
Hence the error parameter vector is the difference
\begin{equation} 
p_0-p=(\vec{\delta g_1}, \vec{\delta g_2},c_1,c_2,
\delta\kappa,\vec{\delta X^0}).
\end{equation} 

We will now prove a  ``weak'' version of regularization that asserts
to first order (the meaning of which is made more precise below)
the difference between the baseline estimate and the true baseline
vector is constant.  Some comments on the second order errors that are
ignored in this analysis are made at the end of this section.

The following nomenclature will be used in the course of establishing
the regularization result.

\noindent{\bf Zeroth order:}
A quantity expressed in meters or radians (e.g. the length of an
interferometer baseline vector.)

\noindent{\bf First order:}
A quantity expressed in terms of
um or  $\mu$rad, or a product of a first order term with a zeroth order term
(e.g., the magnitude of the elastic deformation of the optical truss, s/c
attitude stability, guide star position error.)

\noindent{\bf Second order:} A product of first order
quantities, or a product of a second order term with a zeroth order term.

Let $\vec{\delta b}_s^0$ denote the initial error in the science
baseline vector stemming from the fiducial error term $\vec{\delta
X^0}$.  Also let $\hat\omega_I,~\hat\omega_{II}$ denote the attitude 
rotation vectors in (\ref{eq:om1})-(\ref{eq:om2}) obtained using the
parameter vector $p$ instead of the true vector $p_0$.  Then by taking
differences we obtain
\begin{eqnarray} 
\vec b_s(t;p_0)-\vec b_s(t;p)&=&\vec {\delta b_s^0}+
\vec\omega_I\times\vec {\delta b_s^0} +(\vec
\omega_{II}-\hat\omega_{II})\times \vec {b}_s^{\tt loc}+{1\over 2}(\vec
\omega_I-\hat\omega_I)
\times(\vec \omega_I\times \vec b_s^{\tt loc})+\nonumber\\  & & +~
{1\over 2}\vec \omega_I\times((\vec \omega_I-\hat\omega_I)\times
\vec b_s^{\tt loc})+{1\over 2}(\hat\omega_I-\vec \omega_I)\times[(\vec
\omega_I-\hat\omega_I)\times
\vec b_s^{\tt loc}].
\end{eqnarray}  Retaining first order terms according to the
definitions above,
\begin{equation} 
\vec b_s(t,p_0)-\vec b_s(t;p)=\delta \vec b_s^0 +(\vec
\omega_{II}-\hat\omega_{II})\times \vec b_s^{\tt loc}.
\end{equation}  This can be further simplified to
\begin{equation}
\vec b_s(t,p_0)-\vec b_s(t;p)=\delta
\vec b_s^0+(\vec \omega_{I}-\hat\omega_{I})\times \vec b_s^{\tt loc}
\end{equation}  because to first order $\vec \omega_I=\vec \omega_{II}$
and $\hat\omega_I=\hat\omega_{II}$ since the iteration in (\ref{eq:om2}) is
a second order correction (see definition of $G$ in (\ref{eq:big_g})).
Once we show that $(\vec \omega_{I}-\hat\omega_{I})$ is constant to
first order, we will be done since $\vec b_s^{\tt loc}$ is constant to
zeroth order (its variations are due to elastic motions of the
instrument).

With these definitions the true (second order) attitude equation is
given by
\begin{equation}  T_{\tt ext}\vec \omega+G_{\tt ext}(\vec
\omega)=y+\delta y,
\label{eq:e_dy1}
\end{equation}  while the attitude equation using the parameter vector
$p$ is
\begin{equation}  T_{\tt ext}\vec \omega+G_{\tt ext}(\vec \omega)=y,
\label{eq:e_dy2}
\end{equation}  where $y$ is defined in (\ref{eq:y_martix}) and
\begin{equation}
\delta y=
\left[\matrix{\kappa-\kappa_{est}
\cr\lbrace\cdot\rbrace_I+\lbrace\cdot\rbrace_{II}\cr}\right],
\end{equation}  with
\begin{equation}
\lbrace\cdot\rbrace_I=
\left[\matrix{\langle \vec g_1+\delta \vec g_1,\delta 
   \vec b_g^0\rangle+c_1\cr
   \langle \vec g_2+\delta \vec g_2,
\delta \vec b_g^0\rangle+c_2\cr}\right],
\end{equation}
\begin{equation}
\lbrace\cdot\rbrace_{II}=
\left[\matrix{\langle \vec g_1,\vec \omega\times
\delta \vec b_g^0\rangle+\langle \delta g_1,b_g^{\tt loc}+
\vec \omega\times \vec b_g^{\tt loc}\rangle\cr
     \langle \vec g_2,\vec \omega\times\delta
\vec b_g^0\rangle+\langle\delta \vec g_2, 
\vec b_g^{\tt loc}+\vec \omega\times
\vec b_g^{\tt loc}\rangle\cr}\right].
\end{equation}  Examining these terms it is seen that to first order
$\delta y$ is a constant.

The equations (\ref{eq:e_dy1})--(\ref{eq:e_dy2}) are solved iteratively
to obtain the true and estimated values of $\vec \omega$.  Thus we have
\begin{equation}
\vec \omega_I=T_{\tt ext}^{-1}[y+\delta y],
\end{equation} and
\begin{equation}
\hat\omega_I=T_{\tt ext}^{-1}y.
\end{equation}  Hence,
\begin{equation}
\vec \omega_I-\hat\omega_{I}=T_{\tt ext}^{-1}\delta y.
\end{equation}  Because $T_{\tt ext}$ is a constant matrix to first
order and
$\delta y$ is constant to first order, it follows that
$\vec \omega_I-\hat\omega_I$ is also constant to first order; as was to
be shown.

\begin{figure}
%\vskip-60pt
\centering\epsfig{file=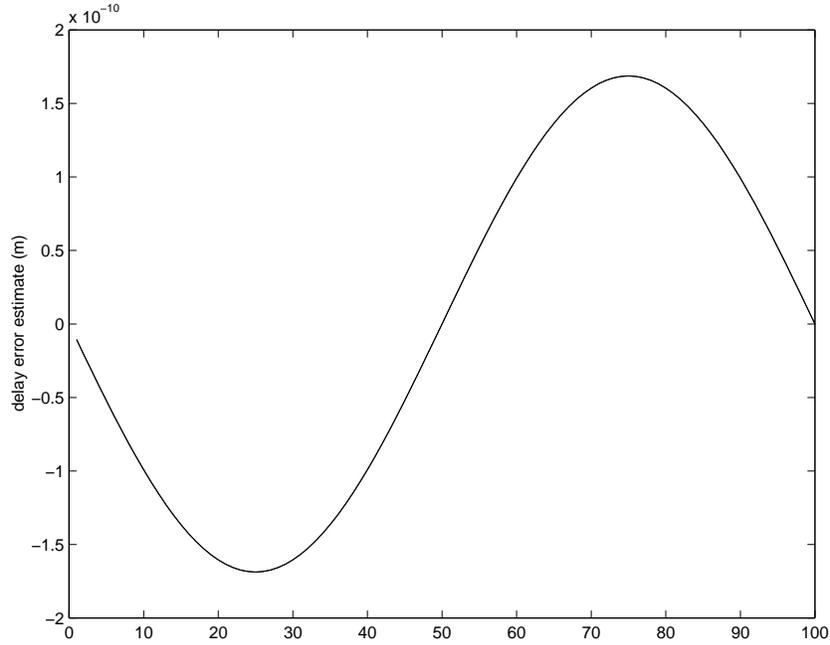,width=11cm}\\%[-70pt]
\caption{Delay error due to absolute metrology and ACS errors.}
\label{fig:fig3.2}
\end{figure}

It is possible to develop very specific forms for the second order error
terms due to these initialization errors, but this is beyond the scope
of the present paper.  We remark that these errors do drive certain
requirements in the SIM astrometric error budget.  For example there is
a relatively strong coupling between the initial fiducial error $\delta
\vec X^0$ and attitude motion that produces a second order delay error
of the form
\begin{equation}  d(t)=[\omega_I(t)\times(\vec{\delta b}_s^0-
\vec{\delta b}_g^0\, \frac{|\vec{b}_s^{\tt loc}|}{|\vec{b}_g^{\tt loc}|}
%|\vec{b}_s^{\tt loc}|/|\vec{b}_g^{\tt loc}|
)]\cdot
\vec{s}.
\end{equation}  
In Figure \ref{fig:fig3.2} we plot this delay regularization
error for a ``typical'' case arising from a single gauge
absolute metrology error of 5 $\mu$m in determining the fiducial
positions and a 2 arcsec single axis rigid body motion generated from a
random rotation direction with a 10 $\mu$rad amplitude sinusoidal motion. 
The plot contains the associated delay error for a typical star within the
field of regard of the science interferometer.  If the abscissa is the
time--axis in seconds, and say a 15 sec observation of the star is made, then
the delay error of the observation would be the average value over that
particular 15 sec period.  The observation period could be at the
\emph{extremes} of the motion, producing a delay error of $\pm 150$ pm, or it
could be centered around zero producing zero delay error. In fact any delay
error could be attained within this envelope. Requirements are set to keep
this envelope small.    

%-------------
%*******************************************************

\section{Summary, conclusions and future plans}
\label{sec:summ}

SIM astrophysical science is extracted from a model set of 
equations that relate the measured optical pathlength delay to the
projection of the interferometer baseline vector onto the star direction
vector.  These equations presuppose that the interferometer baseline is
fixed in inertial space; which it is not.  The main objective of this
paper has been to introduce the reader to the concepts and the
instrumental logic of the SIM astrometric observations, especially as
they relate to the fundamental operation of baseline regularization that
``fixes'' the interferometer in inertial space.  Mathematical arguments
were presented to establish this fundamental principle and a precise
definition of regularization was given.   The underlying nonlinear
system of equations that is the basis for regularization was derived and
numerical methods to solve them were obtained.  A simulation was also
developed incorporating the numerical processing methods of the
instrument observables and  the results were shown to conform with the
theory.

Beyond demonstrating the SIM proof of principle, the regularization
equations presented here also form the kernel of the extensive
instrument error budget.   The linearized version of these equations
are used to determine the propagation of noise from external metrology
measurements and guide star delay measurements to the science delay
error.  This error is determined by the geometry of the optical truss
together with its orientation in inertial space with respect to the
guide star and target star positions. The propagation factors are used
to set requirements on the integration time of observations, single
gauge metrology error, etc.  The regularization equations also reveal
the existence of  a number of second order errors that arise in the
form of products of fiducial motion (elastic and rigid body) and initial
parameter error. Examples of errors of this type were given.

Current work focuses on the mechanisms and effects of variation of the
``constant'' terms in the astrometric delay equations.  Nominally the
appearance of the constant term compensates for the lack of a precise 
internal metrology gauge that measures the absolute distance between the
interferometer aperture fiducials to the beam combiner.  However, there
are a number of instrument errors that are collected into this single
term.  For example as the interferometer observes stars within its field
of regard several optical elements must be translated and rotated. 
These induce non--trivial diffraction effects, metrology gauge error
due to imperfect corner cubes,  reflection phase errors because of a 
changing angle of incidence of the interrogating metrology beams, and
others.  Each of these effects must be played through the delay
regularization equations to ascertain their ultimate effect on delay
error.

%%%%%%%%%%%%%%%%%%%%%%%%%%%%%%%%%%%%%%%%%%%%%%%%%%%%%%%%%%%%% 
\acknowledgments  %>>>> equivalent to \section*{ACKNOWLEDGMENTS}       

This work was performed at the Jet Propulsion Laboratory, California
Institute of Technology, under contract with the National Aeronautics and
Space Administration.

%%%%%%%%%%%%%%%%%%%%%%%%%%%%%%%%%%%%%%%%%%%%%%%%%%%%%%%%%%%%%

%%%%% References %%%%%

%% edit the following to include your own references
%% the entries must be in the order of citation in the manuscript text

\end{document}